\documentclass[slac_one]{revtex4}
\usepackage{graphicx}
\usepackage{fancyhdr}
\begin{document}

\title{{\small{2005 International Linear Collider Workshop - Stanford,
U.S.A.}}\\ 
\vspace{12pt}
Digital Hadron Calorimetry with Glass RPC Active Detectors} 

%

\author{A. Ghezzi, T. Tabarelli de Fatis, G. Tinti}
\affiliation{Universit\`a di Milano Bicocca and INFN, Milano, Italy}
\author{M. Piccolo}
\affiliation{Laboratori Nazionali di Frascati, INFN, Frascati (Roma), Italy}

\begin{abstract}
Glass RPC detectors are an attractive candidate for the active
part of a highly granular digital hadron calorimeter (DHCAL) at the
ILC. A numerical study, based on the GEANT3 simulation package, of
the performance of such a calorimeter is presented in this work.
A simplified model for the RPC response, tuned on real data, is
implemented in the simulation. The reliability of the simulation
is demonstrated by comparison to existing data collected with a large
volume calorimeter prototype exposed to a pion beam in an energy range
from 2 GeV to 10 GeV. In view of an optimization of the readout pitch,
a detailed study of the energy and position resolution at the single
hadron level for different read-out pad dimensions is presented. These
results are then used in a parametric form to obtain a preliminary
estimate of the contribution of DHCAL to the reconstruction of the
energy flow at the ILC detector.

\end{abstract}

\maketitle

\thispagestyle{fancy}


\section{Introduction} 
Common wisdom, based both on experimental results from LEP and from
simulation for higher energy $e^+e^-$ interactions, leads to believe
that the ``energy flow'' technique might provide the best possible
resolution on jet energy, jet direction and jet invariant mass. This
technique requires building very segmented calorimeters both
longitudinally and transversely and leads naturally to the use of
gas detectors, at least in the hadronic part. Digital (track counting)
calorimetry in this framework seems to be the logical choice and is
one of the options proposed for the ILC detector. Out of the possible 
active detectors one could use to implement digital calorimetry
are parallel plate detectors, thanks to the easy of construction, the
lack of privileged directions (in contrast to wire chambers), the
possibility to shape them almost at will and the possibility to cover
large surfaces at low cost. 

An R\&D programme  (CaPiRe experiment) was launched in 2003 aimed at
the design and characterization of RPC detectors based on glass
electrodes for this application. Preliminary results, showing that
glass RPCs can stand the flux of particles expected in the hadron
calorimeter at the ILC, have been reported elsewhere
\cite{LCWS04}. This is instead a report of a numerical 
study of the performance of a Digital Hadron Calorimeter (DHCAL) using 
glass RPCs as active detectors. After a discussion of the basic
features of the calorimeter implemented in our simulation and of a
comparison to existing data 
(Section \ref{Simul}), a detailed study of the energy resolution at
the single hadron level for different read-out configuration is
addressed (Section \ref{Optim}. Finally (Section \ref{FastSim}) an
estimate of the contribution of DHCAL to the reconstruction of the
energy flow at the ILC detector is performed. 

\section{\label{Simul} DHCAL simulation: model parameters and validation}

A standalone simulation of a detector similar in geometry to the
hadron calorimeter proposed for the TESLA detector \cite{TeslaTDR} has
been implemented with the GEANT3 package \cite{GEANT}, which performs
particle tracking and interactions in the detector media. The GEISHA
model is used for hadron cascades. Our DHCAL model 
consists of 39 stainless steel layers 2~cm thick, interleaved  by 38
planes of sensitive elements. The transverse dimensions of the
detector are of little relevance in this study. 

Track counting is performed by Resistive Plate Chambers (RPC) with 
glass electrodes. They are assumed to be housed in a 20~mm gap between 
the stainless steel planes and are described with coarse detail in the
simulation: the active element is a 2 mm gas-filled gap; all the other
elements (the 2 mm thick glass electrodes, the readout plane and the
empty gaps) are modelled as a continuous medium of 1 g/cm$^2$ average 
density. An approximate description of the digitization process has 
been implemented, based on our present knowledge of the RPC
behaviour. The active part of each RPC is segmented in 2 mm $\times$
2~mm elementary cells, matching the typical transverse dimensions of 
a spark in streamer mode. Digital signals are induced on readout pads
if at least one of the elementary cells subtended by it fires. This is
assumed to happen with 95\% efficiency when an ionizing particle
crosses an elementary cell. No induction from cells farther apart from
the pad is included. As discussed hereafter, the transverse pad
dimensions have been varied from $1\ \times\ 1\ \rm{cm}^2$ to $10\ 
\times\ 10\ \rm{cm}^2$. 

The reliability of this simulation of the DHCAL response to hadrons
depends on the description of hadron interactions and on the
assumption made in the description of the digitization at the readout 
stage. This has been studied by comparing our simulation to existing
data collected with a small prototype of the MONOLITH experiment
exposed to a pion beam of energy from 2 GeV to 10 GeV \cite{rpc99}. 
The prototype consisted in 20 iron planes 5 cm thick, instrumented with 
glass RPC readout by 1~cm readout strips on one side of the chambers. 
A part form these differences in the detector layout, accounted for 
in the simulation, the simulation is run under the same assumptions
adopted in the DHCAL description: a digital readout of the fired
strips is considered, without induction on adjacent strips. This is a
rude approximation, as, for this read-out pitch, an average hit
multiplicity of about 1.4 hits/plane has been observed with isolated
tracks. To account for this in the comparison to the MONOLITH
prototype data a 1.4 scale factor on the hit multiplicity predicted by
the simulation is introduced. 

\begin{figure*}[t]
\centering
\includegraphics[width=125mm]{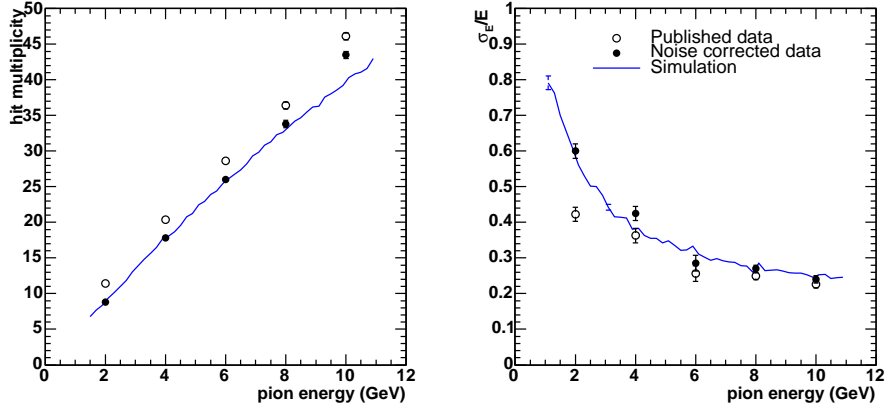}
\caption{Response (left) and energy resolution (right) of the MONOLITH
prototype exposed to a pion beam. Real data (dots) are compared to our
simulation (line). A noise subtraction procedure has been applied to
published data as explained in the text.} 
\label{simAna}
\end{figure*}

As shown in figure \ref{simAna}, a fair agreement between our
simulation and real data is observed, once data are corrected for a
residual noise of about two strips fired at null pion energy. At
variance with real data our simulation seems to show the onset of
saturation around 8-10 GeV (figure 1 left). Yet, the description of
the data in terms of energy resolution is adequate for the purposes of
this study and validates our simulation model in the energy range most
relevant for hadron calorimetry at the ILC, where the spectrum of
neutral hadrons is expected to peak at low energy with an average
energy of around 10~GeV.

\section{\label{Optim} Single hadron energy resolution}

\begin{figure*}[t]
\centering
\includegraphics[width=125mm]{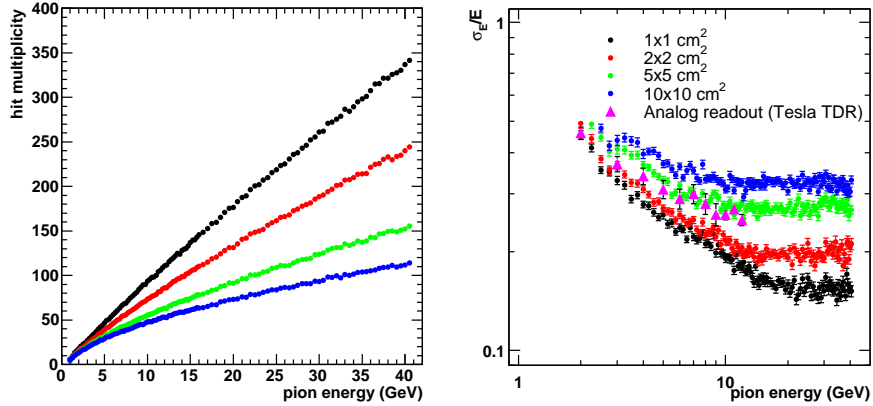}
\caption{Single hadron response and energy resolution for different
  pad read-out dimensions.} \label{lcAna}
\end{figure*}

A full characterization of the response to single hadrons 
of DHCAL with glass RPC detectors in terms of linearity and energy
resolutions has been performed for several different read-out pitch
configurations. Our results for single pions of energy ranging from 2
GeV to 40 GeV and impinging on the calorimeter at normal incidence 
are summarized in figure \ref{lcAna}. The response of the baseline
analog calorimeter proposed in the TESLA TDR \cite{TeslaTDR}, based on
plastic scintillators as active medium, is also shown for comparison.

As expected deviations from linearity grow more important with
increasing pad dimensions. The resolution correspondingly
worsens and the onset of saturation appears at lower energies for
larger pad sizes. Yet, the behaviour is satisfactory in the most
relevant energy region for neutral hadrons at the ILC, at least 
up to 2~cm~$times$~2~cm pad sizes. In this region, DHCAL also presents an
energy resolution better than the expected performance of the baseline
analog calorimeter proposed the TESLA detector \cite{TeslaTDR} (see figure
\ref{lcAna}-right). For $1\ \times\ 1\ \rm{cm}^2$ pads, the resolution  
can be parameterized with a stochastic term of
55\%$/\sqrt{E(\rm{GeV})}$ and a constant term of about 10\%. For the
option with $2\ \times\ 2\ \rm{cm}^2$ pads, the constant term is about
16\%. These results agree with the one presented in an earlier study 
\cite{ammosov}.

\section{\label{FastSim} Energy flow}

A complete understanding of the performance of a RPC based DHCAL
calorimeter and of its contribution to the energy flow
reconstructions requires the development of a full simulation and
event reconstruction in a complex detector, including the tracking
device and the electromagnetic calorimeter. This is beyond the
aim of this preliminary work. Some indications of the DHCAL
contribution to the resolution of the energy flow algorithm can
however be deduced already with a parametric approach. To this end,
the energy flow resolution is parameterized as:
\begin{equation}
\sigma_{E.F}^2 = \sigma_{c.p.}^2 + \sigma_{e.m.}^2 + \sigma_{n.h.}^2 +
\sigma_{confusion}^2,
\end{equation}
where the first three terms describe the energy resolution of the
detector to charged particles, photons and neutral hadrons
respectively and the last term accounts for the finite capability of a
real detector to fully disentangle the energy deposition in the
calorimeters due to charged tracks, because of shower overlaps.
$WW$ and $ZZ$ are generated with the PYTHIA 6.205 code \cite{PYTHIA}
and the energy and momenta of the individual particles have been
smeared according to the expected resolution on their measurement:
\begin{itemize}
\item The energy of the photons has been smeared according to the
  expected resolution of the ECAL calorimeter ($\sigma_{e.m.}/E_{e.m.} =
  15\%/\sqrt{E(\rm{GeV})}$),
\item The energy of neutral hadrons has been smeared according to the
  results of the DHCAL simulation for single hadrons discussed above, 
  including the angular dependence, 
\item The energy of charged particles has been assumed to be perfectly
  reconstructed by the magnetic spectrometer.
\end{itemize}
Events are then clustered in jets and the energy of the jets compared
to the one of the original parton. This ``perfect energy flow''
algorithm, unrealistic indeed, shows a resolution of about
16\%/$\sqrt{E(\rm{GeV})}$. A more realistic energy flow measurement
has been simulated by adding an additional smearing to the jet energy
resolution of 25\%/$\sqrt{E(\rm{GeV})}$ as expected for particle
confusion in a real detector. This last term was estimated in
\cite{Kanen} for a detector configuration with a granularity of
5~cm~$\times$~5~cm in the hadron calorimeter and may be somewhat
overestimated in our case.  

\begin{figure*}[t]
\centering
\includegraphics[width=125mm]{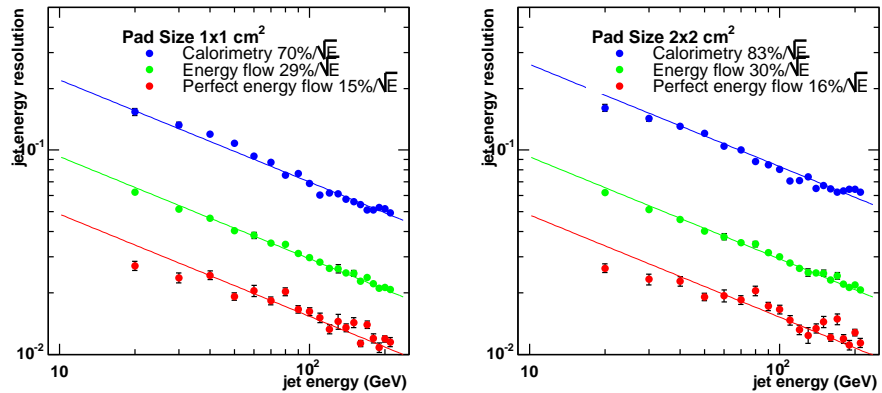}
\caption{Jet energy resolution as a function of the jet energy for $1\
\times\ 1\ \rm{cm}^2$ (left) and  $2\ \times\ 2\ \rm{cm}^2$ (right)
DHCAL granularities. The three set of points have been computed
assuming an energy measurement based solely on the calorimeter
information, to the energy flow approach and to the ``perfect
energy flow'' case (no confusion).} 
\label{HadRes}
\end{figure*}

Our results are shown in figure \ref{HadRes}: a resolution
of about 30\%$/\sqrt{E(\rm{GeV})}$ is observed, dominated by the
contribution of $\sigma_{confusion}$. Within the
approximations discussed, little difference is observed between  $1\
\times\ 1\ \rm{cm}^2$ and  $2\ \times\ 2\ \rm{cm}^2$ granularities. 
indicating that the latter granularity already matches  the desired
performance to accomplish the ILC physics programme. 

\section{Conclusions and outlook}

Glass RPC detectors are attractive candidates as active detectors of
the Digital Hadron Calorimeter at the ILC.
A standalone simulation of the DHCAL response with RPC response has
been developed and validated on existing data.
A fast (parametric) simulation shows that the DHCAL with RPC detectors
with read-out granularities of about  $2\ \times\ 2\ \rm{cm}^2$
matches the target performance for the ILC physics.  
Further studies aimed at the optimization of the detector geometry 
should address the full simulation within the ILC detector and 
the contribution of the highly granular DHCAL to the muon
identification. 

%


\begin{thebibliography}{9}   

\bibitem{TeslaTDR}   
  G.~Alexander et al., ``TESLA Technical Design Report. Part. IV:
  A Detector for TESLA'',
  T. Behnke, S. Bertolucci, R.D. Heuer, R. Settles editors,
  DESY-01-011, DESY-2001-011, DESY-01-011D, DESY-2001-011D,
  DESY-TESLA-2001-23, DESY-TESLA-FEL-2001-05, ECFA-2001-209.

\bibitem{LCWS04}
  A.Calcaterra et al., Nucl Instr. and Meth. A 533 (2004) 154-158; 
  A. Calcaterra et al., ``R\&D on RPC detectors'', International
  Linear Collider Workshop (LCWS 2004), Paris, 2004, Conf. Proc.;
  A. Calcaterra et al., these Proc. 

\bibitem{GEANT} 
  GEANT Detector Description and Simulation Tools, CERN
  Program Library, Long Writeup W5013.

\bibitem{rpc99} 
  C. Gustavino et al., Nucl Instr. and Meth. A 456 (2000) 67

\bibitem{ammosov}
  V.Ammosov, Nucl Instr. and Meth. A 493 (2002) 355

\bibitem{PYTHIA} 
  T. Sj\"ostrand et al., Computer Phys. Commun. 135 (2001) 238 

\bibitem{Kanen} 
  D. Karlen, ``Review od Detector Concepts'',  
  International Workshop on Linear Colliders (LCWS 2002), Korea, 2002,
  Conf. Proc.   


\end{thebibliography}
\end{document}